%% file: preprint.tex
\documentclass[twocolumn]{article}

\usepackage{arxiv}

\usepackage[utf8]{inputenc} 
\usepackage[T1]{fontenc}    
\usepackage{hyperref}       
\usepackage{url}            
\usepackage{booktabs}       
\usepackage{amsfonts}       
\usepackage{nicefrac}       
\usepackage{microtype}      
\usepackage{lipsum}		
\usepackage{graphicx}
\usepackage{natbib}
\usepackage{doi}
\usepackage{multicol}
\usepackage{booktabs}
\usepackage{tabularx}
\usepackage{cuted}
\newcolumntype{Y}{>{\centering\arraybackslash}X}

\graphicspath{{Fig/}}

\title{Terrier: A Deep Learning Repeat Classifier}

\author{ 
    \href{https://orcid.org/0000-0003-1274-6750}{\includegraphics[scale=0.06]{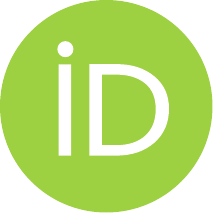}\hspace{1mm}Robert ~Turnbull} \\
    Melbourne Data Analytics Platform\\
    The University of Melbourne\\
    Parkville, VIC 3010 \\
    \texttt{robert.turnbull@unimelb.edu.au} \\
    \And
    \href{https://orcid.org/0000-0001-8756-229X}{\includegraphics[scale=0.06]{orcid.pdf}\hspace{1mm}Neil ~D. ~Young} \\    
    Faculty of Science\\
    The University of Melbourne\\
    Parkville, VIC 3010 \\    
    \And
    \href{https://orcid.org/0000-0003-1157-4897}{\includegraphics[scale=0.06]{orcid.pdf}\hspace{1mm}Edoardo ~Tescari} \\
    Melbourne Data Analytics Platform\\
    The University of Melbourne\\
    Parkville, VIC 3010 \\
    \And
    \href{https://orcid.org/0000-0003-3471-7512}{\includegraphics[scale=0.06]{orcid.pdf}\hspace{1mm}Lee ~F. ~Skerratt} \\
    Faculty of Science\\
    The University of Melbourne\\
    Parkville, VIC 3010 \\    
    \And
    \href{https://orcid.org/0000-0001-5158-5748}{\includegraphics[scale=0.06]{orcid.pdf}\hspace{1mm}Tiffany ~A. ~Kosch} \\
    Faculty of Science\\
    The University of Melbourne\\
    Parkville, VIC 3010 \\
}

\renewcommand{\headeright}{Preprint}
\renewcommand{\undertitle}{Preprint}
\renewcommand{\shorttitle}{Terrier}

\hypersetup{
pdftitle={Terrier},
pdfauthor={Robert ~Turnbull},
}

\begin{document}

\begin{strip}
  \centering 
  \maketitle
  \vskip\baselineskip

\begin{abstract}
\input{abstract}
\end{abstract}

\noindent\makebox[\textwidth]{}
  \vskip\baselineskip

    \keywords{
    Transposable Elements (TEs), Deep Learning, DNA Sequence Classification, Amphibians, Flatworms
}

\end{strip}

\input{content}

\bibliographystyle{unsrtnat}
\bibliography{reference}

\end{document}

%% file: abstract.tex
Repetitive DNA sequences underpin genome architecture and evolutionary processes, yet they remain challenging to classify accurately. Terrier is a deep learning model designed to overcome these challenges by classifying repetitive DNA sequences using a publicly available, curated repeat sequence library trained under the RepeatMasker schema. Poor representation of taxa within repeat databases often limits the classification accuracy and reproducibility of current repeat annotation methods, limiting our understanding of repeat evolution and function. Terrier overcomes these challenges by leveraging deep learning for improved accuracy. Trained on Repbase, which includes over 100,000 repeat families---four times more than Dfam---Terrier maps 97.1\% of Repbase sequences to RepeatMasker categories, offering the most comprehensive classification system available. When benchmarked against DeepTE, TERL, and TEclass2 in model organisms (rice, fruit flies, humans, and mice), Terrier achieved superior accuracy while classifying a broader range of sequences. Further validation in non-model amphibian, flatworm and Northern krill genomes highlights its effectiveness in improving classification in non-model species, facilitating research on repeat-driven evolution, genomic instability, and phenotypic variation.

%% file: content.tex
\section{Introduction}

Modern sequencing approaches have dramatically increased the availability of high-quality reference genomes, helped resolve complex, repetitive regions in these genomes and facilitated the in-depth characterisation and curation of repeats, including transposable elements (TEs) \citep{osmanski2023,rhie2021}. However, repeat classification remains a significant challenge, as most repeat libraries were created from a limited number of model species such as \textit{Drosophila} and \textit{Arabidopsis} \cite{ou2019}. For example, in 2015, 90\% of the Repbase database of repeat families were collected from 134 species \cite{bao2015}. As a result, most repeat classifier tools often fail to classify repeats in divergent taxa \cite{osmanski2023,zuo2023}. 
This limitation hinders the ability to study repeat diversity, evolution, and functional impacts.
Given that repeats constitute up to 85\% of eukaryotic genomes, they play a crucial role in shaping genome size and structure \cite{wells2020}. While many repeats are neutral \cite{platt2018}, others have been linked to phenotypic traits such as evolutionary rate \cite{platt2018}, coloration \cite{hof2016,varga2020}, and fertility \cite{flemr2013}, and some have been implicated in genomic instability and disease \cite{platt2018,senft2021}. 
Comprehensive classification of repetitive elements is therefore essential for understanding genome evolution and function.

A widely-used software package for identifying repeats is RepeatModeler \cite{repeatmodeler}, which integrates three \textit{de novo} repeat discovery tools: RECON \cite{RECON}, RepeatScout \cite{Price2005RepeatScout} and LtrHarvest/Ltr\_retriever \cite{Ellinghaus2008LTRharvest}. These methods infer repeat boundaries and family relationships using sequence similarity and structural features. However, their dependence on existing reference libraries for classification limits their effectiveness in non-model taxa, where repeat sequences may be highly divergent or underrepresented in curated databases. Machine learning-based classifiers, such as DeepTE and TERL (see below), have improved classification accuracy by learning patterns beyond simple sequence similarity. Yet, their performance remains constrained by the size and diversity of available training datasets, leading to reduced effectiveness when applied to species with distinct repeat landscapes. Currently, no available tool can consistently and accurately classify repeat-like elements across a broad range of species, which limits our understanding of conserved and species-specific repeat element evolution.

Here, we introduce Terrier, a deep learning model designed to improve repeat classification across eukaryotes. Trained using the expansive Repbase library \cite{bao2015}, Terrier enhances classification accuracy by leveraging deep learning for fast and accurate prediction of TEs. We validate Terrier's performance against similar packages in four model organisms: rice (\textit{Oryza sativa}), fruit fly (\textit{Drosophila melanogaster}), human (\textit{Homo sapiens}), and mouse (\textit{Mus musculus}). We then explore its effectiveness in non-model species of amphibians, flatworms and northern krill. By expanding the scope of repeat classification, Terrier provides a powerful tool for studying repeat diversity, genome evolution, and repeat-driven phenotypic variation.







\section{Previous Approaches}

\subsection{TEclass}

TEclass is a software package for classifying TE consensus sequences \cite{TEClass}. It represents sequences as frequency vectors of tetramers and pentamers and employs a hierarchical binary classification strategy using support vector machines. First, it determines whether a sequence is a DNA transposon or a retrotransposon; if the latter, it further classifies it as an LTR or non-LTR element, and if non-LTR, it distinguishes between LINEs (long interspersed nuclear elements) and SINEs (short interspersed nuclear elements). To account for variability in TE sequence lengths, separate classifiers are trained for different length bins.

TEclass achieved a sensitivity of over 90\% for DNA transposons but only 75\% when distinguishing between LINEs and SINEs. The authors of TEclass, Abrusán et al., attribute this decrease to error propagation from earlier steps in the pipeline.

\subsection{REPCLASS}

REPCLASS is a Perl-based pipeline for classifying TEs \cite{REPCLASS}. It consists of a homology module which uses the input sequence as a query in a TBlastX search using a reference library \cite{blast}. TBlastX increases sensitivity by translating both DNA sequences into proteins, making it easier to detect conserved protein patterns. A second module looks at the structure of the sequence, while a third module identifies target site duplication. These various modules allow REPCLASS to predict with more categories than TEclass. It requires the WU-BLAST (Washington University BLAST) package \cite{gish2003wublast}, which is no longer maintained.

\subsection{PASTEC}

PASTEC (Pseudo Agent System for Transposable Element Classification) \cite{PASTEC}  was introduced to classify sequences into twelve categories at the order level according to the Wicker hierarchical TE classification system \cite{wicker2007}. It incorporates several methods similar to REPCLASS, including homology-based searches and structural feature detection. Its key innovation is to use hmmer3 \cite{hmmer3} for HMM profile detection, which was useful in cases where the query sequence was not similar enough to sequences in its database.

PASTEC demonstrated higher classification accuracy compared to existing tools. On Repbase update 15.09 (after removing redundant sequences), it had a misclassification rate of only 15.8\%, significantly lower than TEclass (59.7\%) and REPCLASS (33.3\%). However, it classified only 38.2\% of the available sequences, leaving many unclassified due to insufficient evidence.

\subsection{DeepTE}
DeepTE \cite{DeepTE} is a deep-learning-based tool for transposable element (TE) classification. It represents sequences as \textit{k}-mer frequency vectors (ranging from trimers to heptamers) and applies a hierarchical classification approach using a convolutional neural network (CNN) with pooling layers. DeepTE trains eight separate models to perform stepwise classifications, distinguishing TEs from non-TEs and further categorizing them into classes, orders, and superfamilies.

Through systematic evaluation of \textit{k}-mer sizes (\textit{k}=3 to \textit{7}), DeepTE found that heptamers (\textit{k}=7) provided the best precision overall, though the optimal \textit{k}-mer size varied across TE groups. Compared to PASTEC, DeepTE achieved higher sensitivity for most TE categories, detecting more true positives, while PASTEC had fewer false positives in some cases. Moreover, DeepTE was over 18 times faster than PASTEC on a GPU-enabled system.

\subsection{TERL}
TERL (Transposable Elements Representation Learner) \cite{TERL} is another deep learning model using convolutional neural networks. TERL first represents the DNA sequences using one-hot encoding. These representations are passed through a succession of three convolution and pooling layers and then given to three fully connected layers to make the classification. It is able to classify TEs into nine orders and 29 superfamilies using the Wicker classification system \cite{wicker2007}. It achieved a macro mean F1 score of 85.8\% for a dataset derived from Repbase 23.10. The authors of TERL compared it with PASTEC and TEclass across multiple experiments. It consistently outperformed TEclass in all metrics. Compared to PASTEC, TERL achieved similar performance in several cases, but PASTEC showed higher precision in some categories. However, TERL was orders of magnitude faster than PASTEC, as PASTEC relies on computationally expensive homology-based searches, while TERL uses raw sequence data and CNN-based feature extraction, allowing for efficient GPU computation.

\subsection{TEclass2}
Bickmann et al. recently developed a deep learning model for classifying repeated sequences called TEclass2 \cite{Bickmann2023.10.13.562246}. This model uses a sliding window to produce \textit{k}-mers, then uses a transformer encoder \cite{transformer} before making the classification. They used the curated and non-curated Dfam 3.7 database, an open collection of transposable element families \cite{Dfam}, with Repbase version 18 as a dataset and used sixteen superfamilies from the Wicker classification system \cite{wicker2007} as the prediction categories. They chose the uncurated Dfam database because it is vastly larger than the curated version, while acknowledging that it may produce incorrect results due to misclassified repeat families. They achieved an average accuracy of 79\% with a macro-averaged accuracy of 72\% on the roughly 132,000 sequences in the validation dataset. TEclass2 outperformed TERL and DeepTE in terms of accuracy but typically fewer TE reached the threshold for classification (see Table \ref{table:testresults}). The software is available but does not include the trained weights. A trained version of the model is available through a web interface (\url{https://bioinformatics.uni-muenster.de/tools/teclass2/index.pl}).
 
\section{Data}

\begin{figure}[htpb!]
\centerline{
\includegraphics[width=0.4\textwidth,alt={A hierarchical tree in textual format showing the 9 RepeatMasker Types and 44 SubTypes used by Terrier. It also shows the number of sequences in Repbase which are mapped to each node in the tree.}]{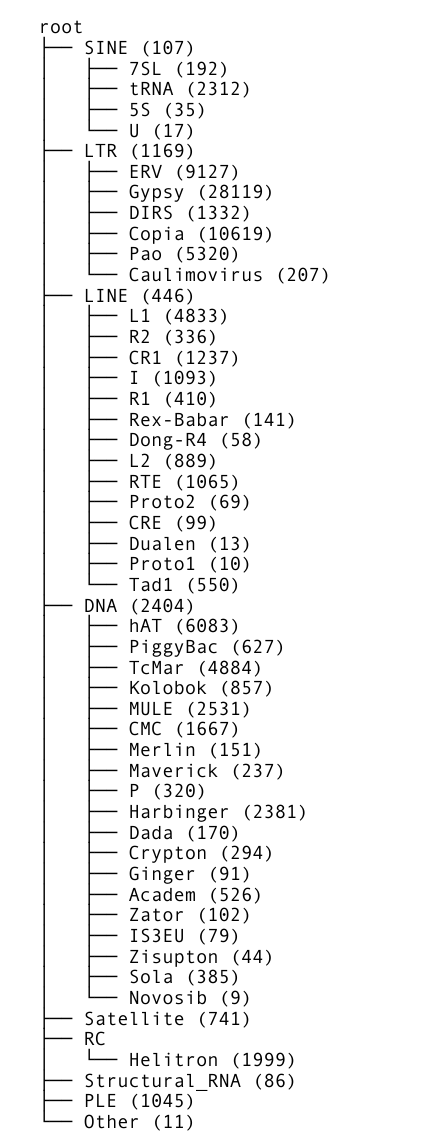}
}
        \caption{The tree used to classify the repeat families, showing the number of sequences in Repbase mapped to each node.}
        \label{fig:tree}
\end{figure}

Two databases are currently available for repeat classification of any species, Repbase \cite{bao2015} and Dfam \cite{Dfam}. While both are widely used for training and classification purposes, Repbase contains manually curated entries for many species, including non-model organisms. Whereas Dfam uses HMM-based models and includes many \textit{de novo} predicted families that are yet to be subject to manual curation. We used the 29.10 release of the Repbase database for training and cross-validation. This release has just over 100,000 repeat families, approximately four times the number of families as Dfam version 3.8. A challenge with this approach is that Repbase uses a different classification system than RepeatMasker. To ensure compatibility, outputs must be mapped to RepeatMasker categories. Smit and Hubley (2018) produced a version of Repbase aligned with the RepeatMasker classification system, and its metadata is included in the RepeatMasker GitHub repository (\url{https://github.com/rmhubley/RepeatMasker}). However, this edition of Repbase only has 40,000 families, far fewer than more recent Repbase releases. To address this, the Terrier package includes a translation layer that takes 156 Repbase categories and maps them to 9 RepeatMasker Types and 44 SubTypes corresponding to Order and Superfamily classifications to enhance compatibility across annotation tools. This is the largest number of prediction categories among the software packages surveyed above. These categories allow for the translation of 97,529 consensus sequences (97.1\%) from Repbase 29.10. The full listing of the number of families in each category is provided in Fig. \ref{fig:tree}. 

We divided the families of the dataset into five stratified cross-validation partitions so that each partition would have the same proportion of each Order and Superfamily.

The consensus sequences of the dataset were converted to the SeqBank format (\url{https://github.com/rbturnbull/seqbank}) for efficient retrieval of sequence data in binary format for training. 

The scripts for performing the preprocessing of Repbase are included in the Terrier package with step-by-step instructions included in the documentation.

\section{Methods}\label{methods}

\begin{figure}[!tpb]
\centerline{\includegraphics[width=0.5\textwidth,alt={A figure displaying the multiple layers of the neural network architecture used by Terrier. It starts at the bottom with the input sequence which is embedded and goes through a series of n convolutional, dropout and pooling layers, leading to the final prediction of the Order and Superfamily categories.}]{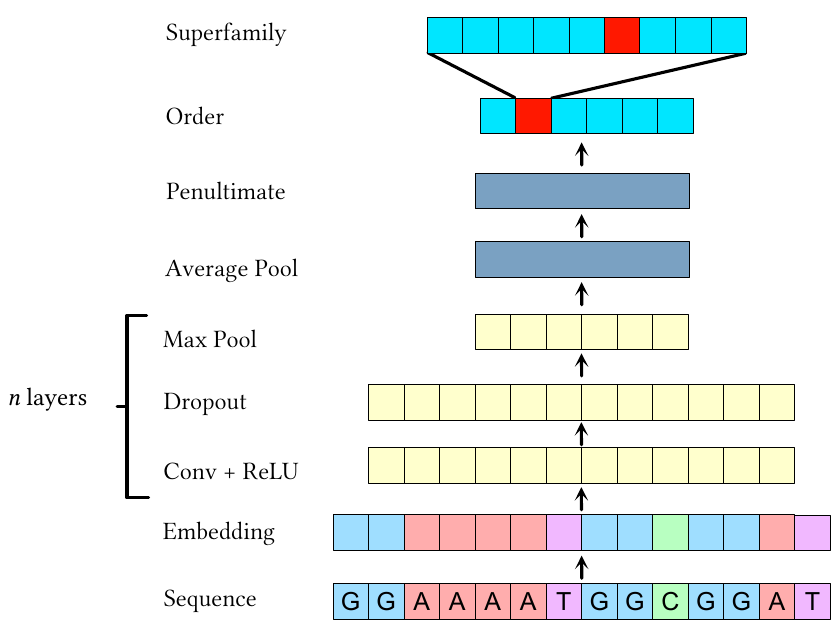}}
        \caption{The Terrier neural network architecture.}
        \label{fig:model}
\end{figure}

To predict the repeat classification from DNA sequences, we employed a neural network architecture based on TorchApp (\url{https://github.com/rbturnbull/torchapp}) and the DNA sequence classifier Corgi (\url{https://github.com/rbturnbull/corgi}). It uses a simple convolutional architecture and was trained to hierarchically predict the repeat Order and Superfamily (Fig. \ref{fig:model}). The model takes the DNA sequence and embeds each type of nucleotide base into a vector space of size $e$. These vectors are then processed with a series $n$ layers which perform a convolution with kernel size $k$, followed by a Rectified Linear Unit (ReLU) activation function, a dropout layer with probability $d$ followed by a max pooling operation which reduces the length of the sequence by a factor of 2. The first convolutional layer has $f$ features and this number of features is increased by a growth factor $g$ at each successive layer. The result of these convolutional layers is averaged globally and this is given to a penultimate linear layer of size $p$ with ReLU activation before going to a hierarchical prediction layer provided by the HierarchicalSoftmax package (\url{https://github.com/rbturnbull/hierarchicalsoftmax}). This hierarchical prediction layer has outputs for each Order and Superfamily and calculates the loss for the ground truth using cross entropy. 
The weighting of the Superfamily component of the loss relative to the Order component is controlled by the hyperparameter $\phi$.
The hierarchical model outputs Superfamily predictions only if their probability surpasses a user-defined threshold; otherwise, it defaults to the Order level.
It can output a CSV with the probabilities of each category as well as the sequences in FASTA format with the classification written after the sequence ID in the header, ready for downstream analysis.

\subsection{Training Procedure and Cross Validation}

\begin{figure}[!tpb]
\centerline{\includegraphics[width=0.5\textwidth,alt={Box plots showing the accuracy over the five cross-validation folds for the Order prediction and the Superfamily prediction. The Order prediction achieves a mean accuracy of 95.2\% and the Superfamily 94.0\%.}]{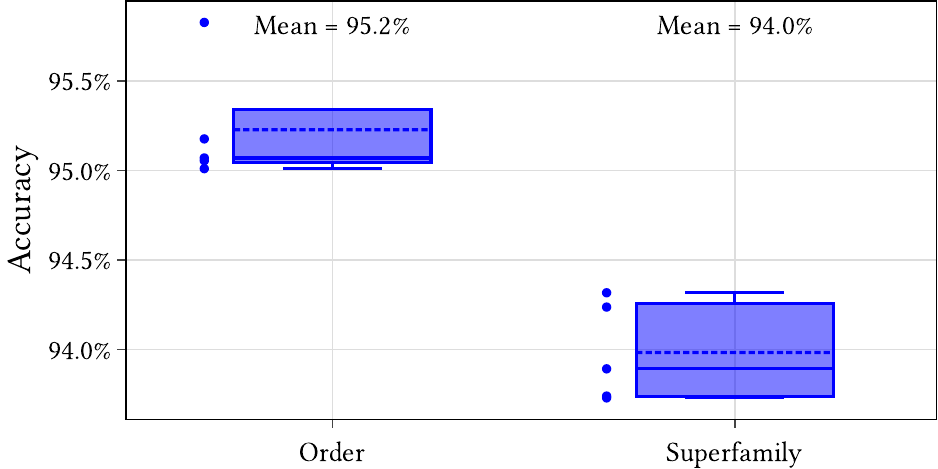}}
        \caption{Results for the five cross-validation folds.}
        \label{fig:xval}
\end{figure}

\begin{figure}[!tpb]
\centerline{\includegraphics[width=0.5\textwidth,alt={The confusion matrix for the concatenated predictions on the five cross-validation sets at the nine categories at the Order level. The total number for each correct prediction is given in the diagonal of the matrix and the incorrect predictions on the off-diagonal entries.}]{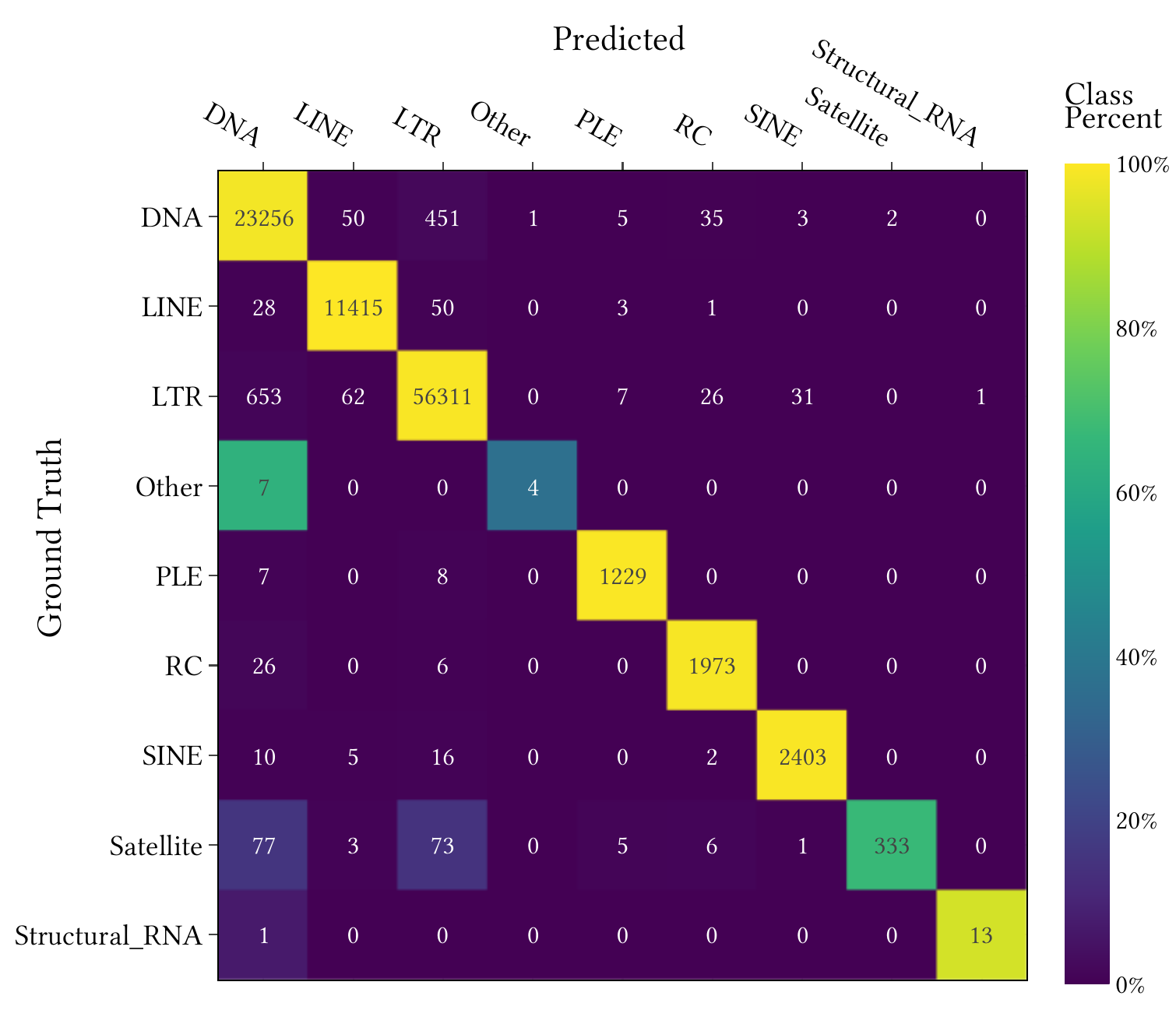}}
        \caption{The confusion matrix for the concatenated predictions on the five cross-validation sets at the Order level.}
        \label{fig:confusion}
\end{figure}

The models were trained for one hundred epochs with a batch size of 32 using the Adam optimization method \citep{adam}. The learning rate was scheduled according to the `1cycle' policy \citep{smith2018} with the peak learning rate set to $10^{-3}$. The hyperparameters were tuned on the first validation partition using the Optuna hyperparameter optimization library \citep{optuna2019}. The number of features in the first convolutional layer $f$ was scaled to constrain the total number of multiply and accumulate (MACC) operations to approximately $2 \times 10^{10}$. The Superfamily accuracy was used as the optimization criterion for twenty runs. The hyperparameters for the optimal training run are displayed in table \ref{table:hyperparameters}. These were used for training models on the four remaining cross-validation partitions and the Type and Superfamily accuracies are displayed in Fig. \ref{fig:xval}. This achieved a mean accuracy across the five cross-validation folds at the Order level of 95.2\% and at the Superfamily level of 94.0\%. The results are comparable across the five cross-validation folds, meaning that the hyperparameter tuning did not overfit to the first validation partition. A confusion matrix for the Order level, found by concatenating the predictions on the five validation sets is displayed in Fig. \ref{fig:confusion}. The final model was trained on the entire training dataset. Steps for reproducing the training of the final model are provided in the documentation.

\begin{table}[hbtp!]
    \centering
    \begin{tabularx}{\linewidth}{Xr}
        \toprule
        \textbf{Hyperparameter} & \textbf{Value} \\
        \midrule
        Embedding Size $e$ & 18 \\        
        CNN Layer $n$ & 4 \\
        Layer Growth Factor $g$ & 1.96 \\
        Dropout $d$ & 0.248 \\        
        Kernel Size $k$ & 7 \\
        Penultimate Layer Size $p$ & 1953 \\
        Superfamily loss weighting $\phi$ & 1.02 \\
        \bottomrule
    \end{tabularx}
    \caption{Optimized hyperparameters obtained from 20 tuning runs on the first validation partition.}
    \label{table:hyperparameters}
\end{table}

\section{Test Results}
\label{test-data}

\begin{figure*}[htpb]
    \centerline{\includegraphics[width=\textwidth,alt={Four scatter plots showing the Superfamily classification accuracy versus proportion classified. The four plots correspond with the four datasets: Rice, Fruit Fly, Human and Mouse. On each plot are the results for each of the software packages.}]{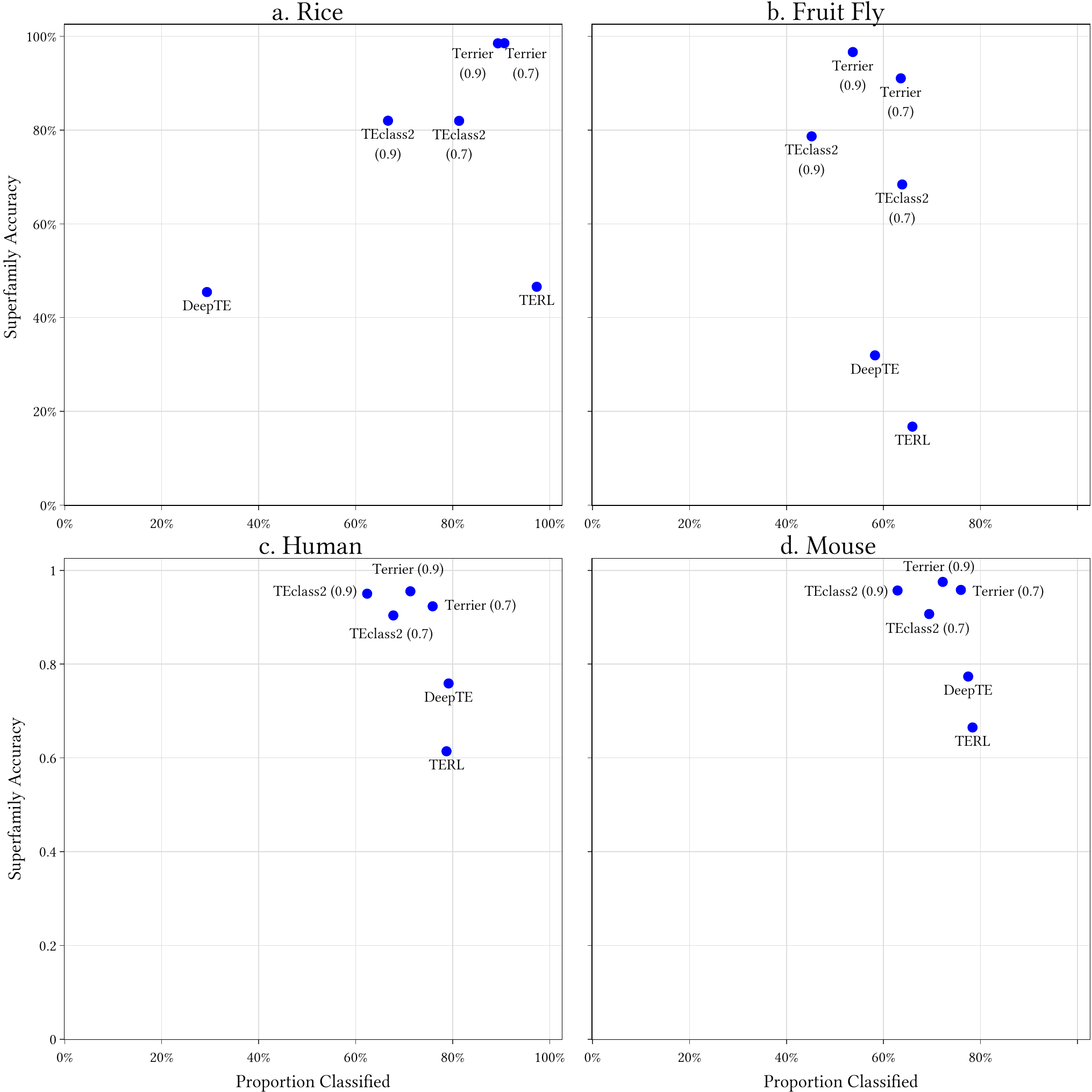}}
        \caption{Superfamily classification accuracy versus proportion classified across four software packages. Preferred results are in the top right corner.}
        \label{fig:testresults}
\end{figure*}

\begin{figure*}[htpb]
    \centerline{\includegraphics[width=\textwidth,alt={Two confusion matrices for Terrier on the Fruit Fly and Rice test datasets using a threshold of 0.9 at the Superfamily level. The two confusion matrices correspond with the Rice and Fruit Fly datasets.}]{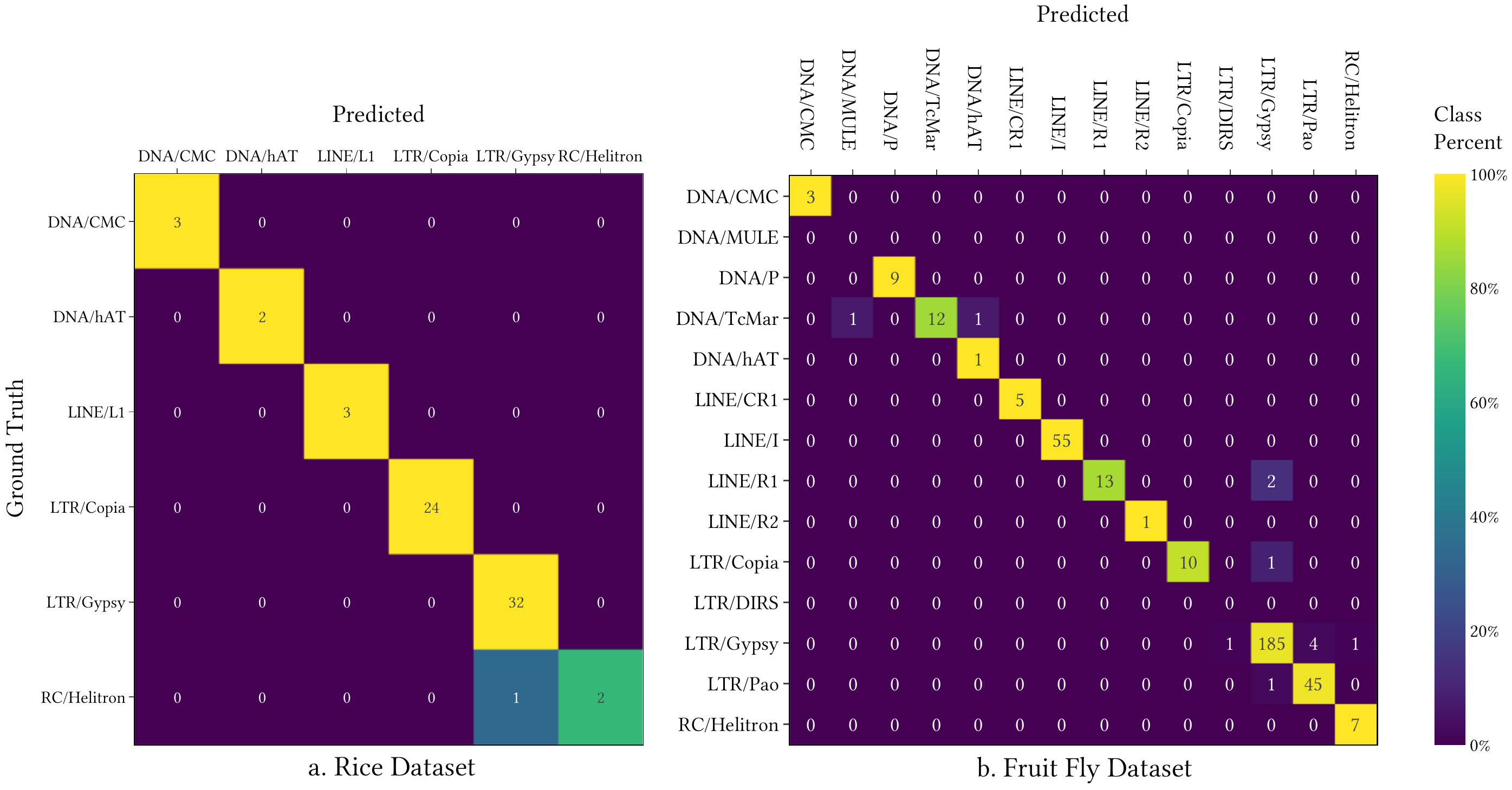}}
        \caption{Confusion matrices for Terrier on the Fruit Fly and Rice test datasets using a threshold of 0.9. Predictions at the `Superfamily' level. Confusion matrices for Terrier on the larger Human and Mouse datasets are available on the Terrier GitHub repository and online documentation.}
        \label{fig:testconfusion}
\end{figure*}

For testing Terrier, we used the rice and fruit fly repeat datasets provided in \cite{Bickmann2023.10.13.562246}. We also use the human and mouse repeat libraries available from msRepDB \citep{msRepDB}. We compared the performance of Terrier with DeepTE, TERL and TEclass2 using the raw results of these applications on the two datasets, which are provided in the GitHub repository for TEclass2 (\url{https://github.com/IOB-Muenster/TEclass2/}). The tool that we used for this comparison is included in the Terrier application. The TEclass2 results use threshold values of 0.7 and 0.9 for the Superfamily predictions. We present results with the same threshold values for Terrier. The results are presented in Table \ref{table:testresults} and plotted in Fig. \ref{fig:testresults}. The confusion matrices for Terrier with a threshold of 0.9 on the rice and fruit fly datasets are shown in Fig. \ref{fig:testconfusion}. The confusion matrices for the other software packages and datasets are displayed in the Terrier documentation along with the steps to reproduce these results.

\begin{table*}[hbtp!]
\centering

\begin{tabularx}{\linewidth}{l *{4}{>{\centering\arraybackslash}X}| *{4}{>{\centering\arraybackslash}X}}
\toprule
 & \multicolumn{4}{c|}{Rice} & \multicolumn{4}{c}{Fruit Fly} \\ 
 & \multicolumn{2}{c}{Order} & \multicolumn{2}{c|}{Superfamily} & \multicolumn{2}{c}{Order} & \multicolumn{2}{c}{Superfamily} \\ 

Software &  Classified & Accuracy & Classified & Accuracy & Classified & Accuracy & Classified & Accuracy \\
\midrule
DeepTE & 78.7\% & 52.5\% & 29.3\% & 45.5\% & \textbf{87.6\%} & 34.9\% & 58.2\% & 31.9\% \\
TERL & \textbf{97.3\%} & 63.0\% & \textbf{97.3\%} & 46.6\% & 73.4\% & 38.8\% & \textbf{66.0\%} & 16.7\% \\
TEclass2 (0.7) & 81.3\% & 86.9\% & 81.3\% & 82.0\% & 64.0\% & 78.6\% & 63.9\% & 68.4\% \\
TEclass2 (0.9) & 66.7\% & 86.0\% & 66.7\% & 82.0\% & 45.3\% & 86.4\% & 45.2\% & 78.7\% \\
Terrier (0.7) & 94.7\% & 94.4\% & 90.7\% & \textbf{98.5\%} & 80.8\% & 87.9\% & 63.6\% & 91.0\% \\
Terrier (0.9) & 90.7\% & \textbf{97.1\%} & 89.3\% & \textbf{98.5\%} & 65.4\% & \textbf{94.5\%} & 53.7\% & \textbf{96.6\%} \\
\midrule
 & \multicolumn{4}{c|}{Human} & \multicolumn{4}{c}{Mouse} \\ 
 & \multicolumn{2}{c}{Order} & \multicolumn{2}{c|}{Superfamily} & \multicolumn{2}{c}{Order} & \multicolumn{2}{c}{Superfamily} \\ 

Software &  Classified & Accuracy & Classified & Accuracy & Classified & Accuracy & Classified & Accuracy \\
\midrule
DeepTE & 90.8\% & 70.1\% & \textbf{79.2\%} & 75.9\% & 89.2\% & 73.0\% & 77.5\% & 77.4\% \\
TERL & \textbf{91.9\%} & 69.4\% & 78.7\% & 61.4\% & \textbf{90.6\%} & 71.1\% & \textbf{78.4\%} & 66.5\% \\
TEclass2 (0.7) & 81.5\% & 88.9\% & 67.8\% & 90.4\% & 78.9\% & 89.0\% & 69.4\% & 90.7\% \\
TEclass2 (0.9) & 73.8\% & 92.0\% & 62.4\% & 95.0\% & 70.5\% & 92.8\% & 62.9\% & 95.7\% \\
Terrier (0.7) & 85.3\% & 89.5\% & 75.9\% & 92.3\% & 84.6\% & 92.1\% & 75.9\% & 95.8\% \\
Terrier (0.9) & 78.0\% & \textbf{94.4\%} & 71.3\% & \textbf{95.5\%} & 78.5\% & \textbf{95.6\%} & 72.2\% & \textbf{97.5\%} \\

\bottomrule
\end{tabularx}
\caption{Comparison of classification accuracy across test datasets. The highest values per column are bolded.}
\label{table:testresults}
\end{table*}

The following test datasets were processed using Terrier with one NVIDIA A100 GPU with two CPUs (two Intel(R) Xeon(R) Gold 6448H). All timings refer to total wall time, including loading the model.

\subsection{Rice}

The rice dataset contains 75 TE models. Terrier classified it in 7.2s. With a threshold of 0.7, 71 models (94.4\%) were classified at the `Order' level with 94.4\% accuracy, outperforming all other tools. Raising the threshold to 0.9 slightly reduced classified sequences (68) but improved accuracy to 97.1\%. At the `Superfamily' level, Terrier classified 68 sequences with 98.5\% accuracy---substantially higher than TEclass2’s best of 82.0\%. TERL was able to classify more families but the accuracy was substantially.

\subsection{Fruit Fly}

This dataset includes 667 TE models, classified in 10.7s. At a 0.7 threshold, Terrier achieved 80.8\% classification at the `Order' level with 87.9\% accuracy. A threshold of 0.9 raised accuracy to 94.5\%, with a drop in classified sequences to 65.4\%. At the `Superfamily' level, Terrier reached 91.0\% accuracy at 0.7 and 96.6\% at 0.9---both outperforming TEclass2 while classifying similar numbers of sequences.

\subsection{Human}

The human dataset has 1613 TE models, classified in 17.4s. At 0.7, Terrier classified 85.3\% of models at the `Order' level with 89.5\% accuracy; increasing the threshold to 0.9 raised accuracy to 94.4\% (78\% classified). At the `Superfamily' level, Terrier showed slightly higher accuracy than TEclass2 at the equivalent thresholds, while classifying more models.

\subsection{Mouse}

The mouse dataset has 1779 models and was classified in 18.0s. At a threshold of 0.9, it achieved the highest `Order' accuracy (95.6\%) while maintaining broad coverage (78.5\%). At the `Superfamily' level, Terrier outperformed TEclass2 at the equivalent thresholds, again with broader coverage.

\begin{figure*}[htpb]
    \centerline{\includegraphics[width=\textwidth,alt={Scatter plots showing the computation time for running Terrier, DeepTE and TERL against the filesize of the input data. The left scatter plot shows timing on just CPUs. The right scatter plots show timings using a GPU. All software packages scale linearly according to filesize. DeepTE and TERL are faster than Terrier using just CPUs but when using a GPU, Terrier is faster than DeepTE whilst still being slower than TERL.}]{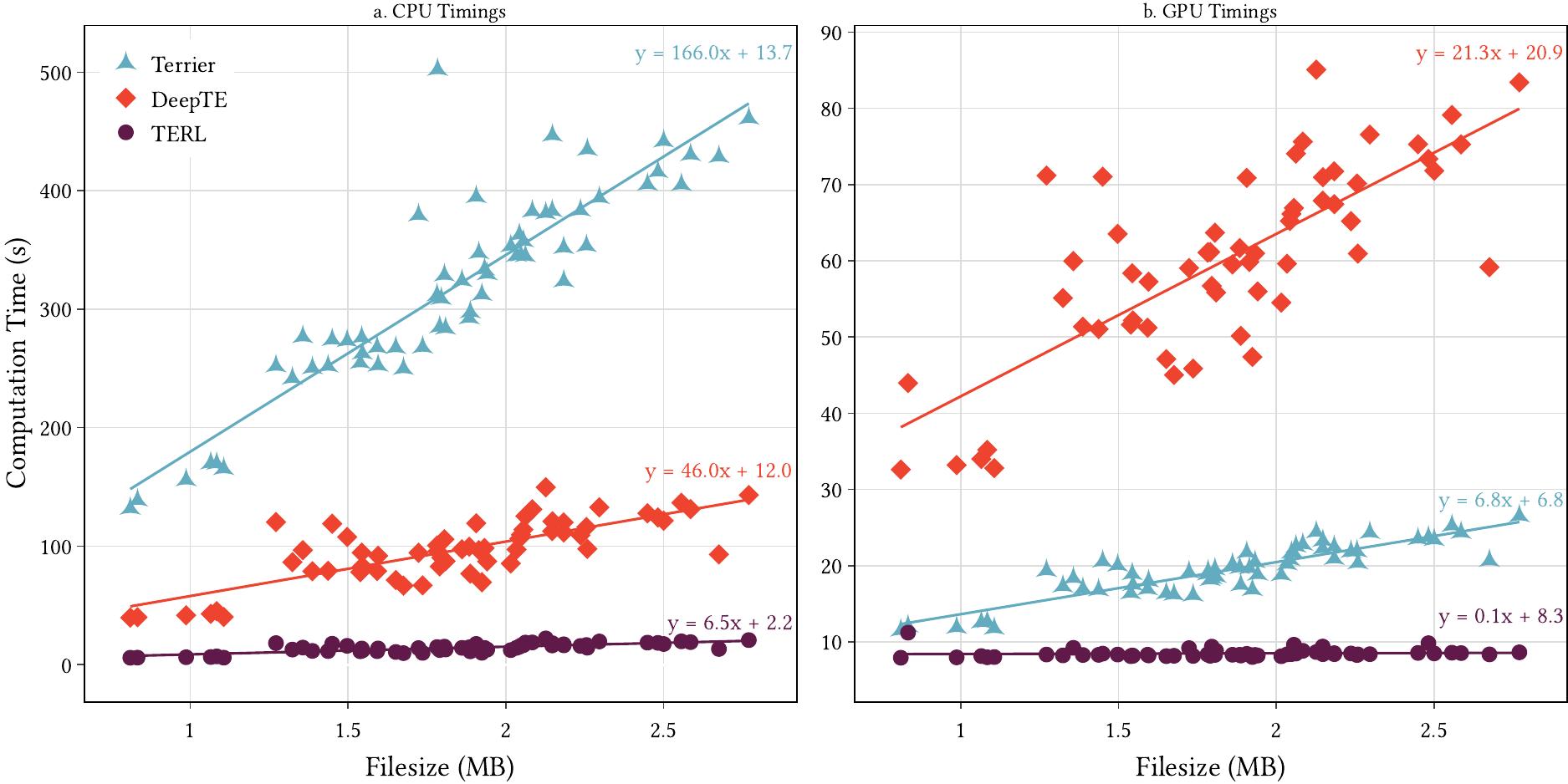}}
        \caption{Computation time for running Terrier, TERL and DeepTE on flatworm and amphibian TE datasets. The filesize refers to the uncompressed FASTA input files in megabytes. A linear trendlines are shown with the equation written on the right.}
        \label{fig:timings}
\end{figure*}

\begin{figure*}[htpb]
    \centerline{\includegraphics[width=\textwidth,alt={Bar charts comparing the results from RepeatModeler (left) and Terrier (right) on the experimental data of 8 flatworms and 51 amphibians. The bar charts visualise the number of DNA, LINE, LTR, Other and Unknown classifications from both software packages. The number of Unknown predictions in the Terrier results is less than the corresponding results from RepeatModeler.}]{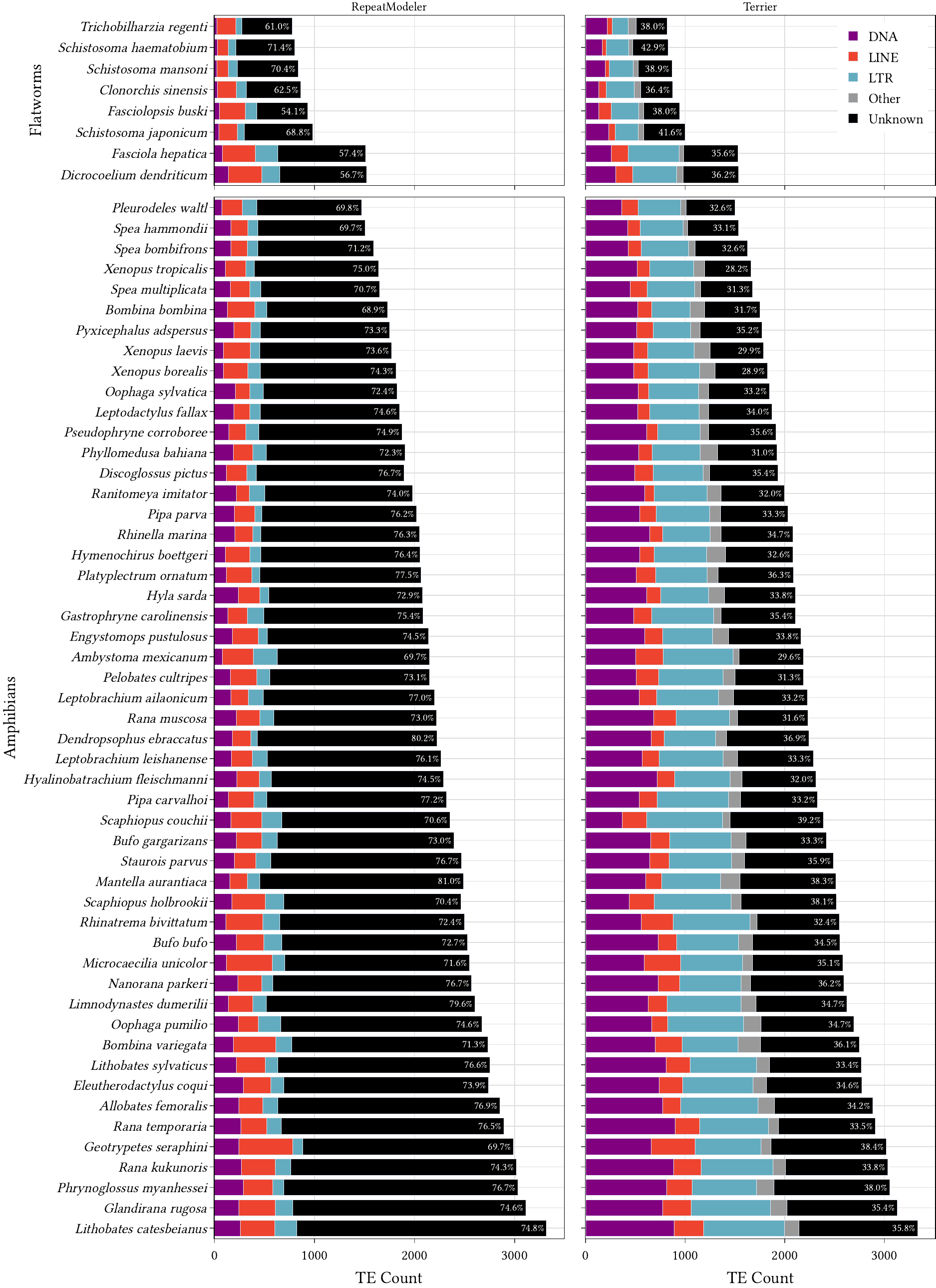}}
        \caption{Comparison between results from RepeatModeler (left) and Terrier (right) on the experimental data of 8 flatworms and 51 amphibians. The percentage of sequences classified as `Unknown' is labeled for each species.}
        \label{fig:application}
\end{figure*}

\section{Experimental Data}

To assess Terrier's performance, we applied it to experimental data of TE libraries from 8 flatworm species and 51 amphibians.

We present the overall computation time for running Terrier, TERL and DeepTE on these input files. We tested timings on an GPU (with two CPUs) and then with two CPUs with no GPU (Fig. \ref{fig:timings}). In all cases, file size was linearly correlated with computational time. Using the CPUs, Terrier scaled at approximately 166s/MB. DeepTE performed faster at around 46s/MB and TERL was faster still at 6.5s/MB. Terrier performed many times faster with a GPU relative to using the CPUs, scaling at 6.8s/MB. This was faster than DeepTE at 21.3s/MB but still not as fast as TERL which was extremely fast at 0.1s/MB. These timings show that Terrier is able to produce classifications for even large genomes within a reasonable timeframe, especially when a GPU is available.

In Fig. \ref{fig:application} we show the Order classifications from both RepeatModeler and Terrier at a threshold of 0.7. The RepeatModeler output gives 73.5\% as `Unknown' whereas the Terrier classifications, this number is reduced to only 34.4\% `Unknown'. The substantial reduction in the number of unclassified repeat families is consistent across amphibians and flatworms of different species and genome sizes.

One of the flatworm species, \textit{Schistosoma mansoni}, has 21 TE sequences annotated in the NCBI database. Terrier correctly classified 20 of these sequences as LTRs. The remaining non-LTR sequence belongs to the SR2 subfamily and lacks a terminal repeat region. This was classified by Terrier as a DNA transposon with probability 0.72.

These experimental results across a wide range of flatworm and amphibian species demonstrate that Terrier can efficiently classify transposable elements in large, repeat-rich genomes. Terrier considerably reduces the number of unclassified repeat families compared to RepeatModeler alone, enabling more comprehensive downstream analysis.

\subsection{Northern Krill}

\begin{figure}[htpb]
    \centerline{\includegraphics[width=0.5\textwidth,alt={Bar charts showing the number of unknown repeats classified by Terrier for the northern krill at probability thresholds of 0.7 and 0.9. The stricter threshold of 0.9 results in fewer repeats being classified.}]{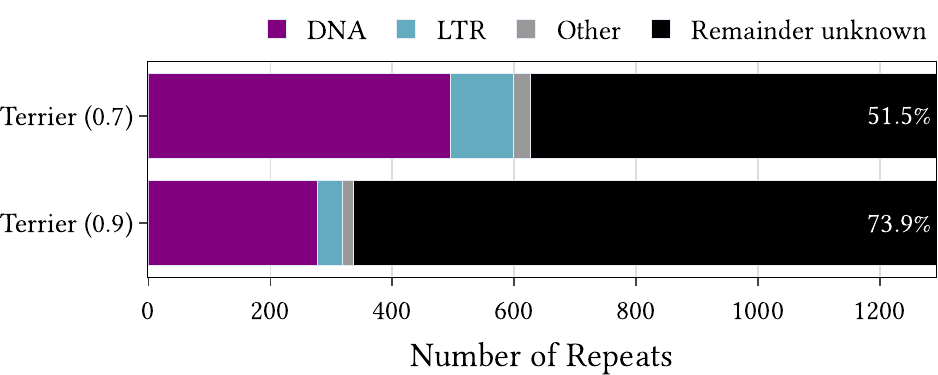}}
        \caption{Extra classifications by Terrier at different probability threshold of previously unclassified repeat families from northern krill. The percentage of sequences remaining unknown is labeled for each threshold.}
        \label{fig:terrier-krill-now-known}
\end{figure}

To further demonstrate use-cases for Terrier, we applied it classifying unknown TE libraries in northern krill (\textit{Meganyctiphanes norvegica})---a large genome of more than 19 Gb. Unneberg et al. \cite{unneberg2024} recently found that repeats account for 74\% of the genome. They released a library of 10909 distinct repeat sequences, of which 1292 (11.8\%) were unclassified. We used Terrier to classify these unclassified sequences (Fig. \ref{fig:terrier-krill-now-known}). With the default threshold of 0.7, Terrier classified 626 (48.5\%) to at least the Order level with 162 to the Superfamily level. These repeat families correspond to more than 10 million individual repeats summing to approximately 2.5 Gb. With the more restrictive threshold of 0.9, Terrier classified 337 (26.1\%) repeat families to at least the Order level, with 70 to the Superfamily level, corresponding to almost 6 million individual repeats summing to almost 1.5 Gb.

\section{Conclusion}\label{conclusion}

In this study, we introduce Terrier, a comprehensive software package designed for the precise classification of repeats from DNA sequences. Terrier distinguishes itself by incorporating a significantly greater number of prediction categories than other comparable methods, providing a finer level of detail for repeat classification. This expanded classification system enables more accurate and comprehensive assessments, especially when dealing with complex, diverse, or poorly characterized genomic regions.

Leveraging deep learning techniques, Terrier not only improves classification accuracy but also delivers exceptionally fast results with minimal computational overhead. Its ability to process large datasets efficiently is particularly evident when running on GPUs, making it an invaluable tool for both large-scale studies and routine use in genomic research.

Terrier substantially reduced the number of unclassified repeats for experimental data of flatworms, amphibians and northern krill, demonstrating its effectiveness for large, highly repetitive genomes.

Compared to other state-of-the-art deep learning classifiers, Terrier demonstrates superior performance in terms of both classification accuracy and the breadth of sequences it can classify. The tool’s high precision, combined with its broad applicability across a range of species and repeat types, positions it as a leading tool for repeat classification. Terrier integrates seamlessly into existing repeat annotation workflows by running between the standard tools RepeatModeler and RepeatMasker. These capabilities make Terrier an essential tool for researchers studying repeat-driven evolution, genomic instability, and other areas where a detailed understanding of repeats is crucial. By advancing the field of repeat classification, Terrier enables more accurate genomic annotations and supports the exploration of complex genomic features in both model and non-model organisms.

\section{Data availability}\label{availability}

The software is available under an Apache 2.0 Open Source License and can be downloaded and installed from GitHub (\url{https://github.com/rbturnbull/terrier}) and from the Python Package Index (\url{https://pypi.org/project/bio-terrier/}). Weights for the model are available with the 0.2.0 release of the software and this is automatically downloaded when first using the model. We also provide a notebook which can be launched in Google Colab to run Terrier on a cloud GPU and to download the results.

The experimental amphibian and flatworm data underlying this article are available in FigShare, at \url{https://doi.org/10.26188/28578992}. New repeat classifications for northern krill are available in FigShare, at \url{https://doi.org/10.26188/29226188}.

\section{Competing interests}
No competing interest is declared.

\section{Author contributions statement}

R.T. conceived and wrote the Terrier software and performed the training. R.T., N.Y. and T.K. prepared the training dataset. R.T. and E.T. ran the software packages on the test datasets. N.Y. prepared the flatworm experimental data. T.K. and E.T. prepared the amphibian experimental data. All reviewed the manuscript.

\section{Key Points}

\begin{itemize}
    \item Terrier is a deep learning model trained on the extensive Repbase library, designed to improve repeat classification accuracy across a broad range of eukaryotic species.
    \item Terrier outperforms existing tools such as DeepTE, TERL, and TEclass2 on test datasets from rice, fruit flies, humans, and mice.
    \item We demonstrate its effectiveness on experimental data from non-model organisms of flatworms and amphibians. These datasets are available on FigShare under a Creative Commons open access license.
    \item Terrier is available under the Apache 2.0 open source license, along with trained model weights. Documentation includes instructions to reproduce the results.
\end{itemize}

\section{Funding}
This project benefited from Australian Research Council grants FT190100462 and LP200301370 awarded to L.F.S.

\section{Acknowledgments}
This research was supported by The University of Melbourne’s Research Computing Services. We acknowledge the help of Priyanka Pillai, Swetha Gopikumar Sreeja and Rafsan Al Mamun.

\clearpage